\documentclass[lettersize,journal]{IEEEtran}
\usepackage{amsmath,amsfonts}
\usepackage{algorithmic}
\usepackage{algorithm}
\usepackage{array}
\usepackage[caption=false,font=normalsize,labelfont=sf,textfont=sf]{subfig}
\usepackage{textcomp}
\usepackage{stfloats}
\usepackage{url}
\usepackage{verbatim}
\usepackage{graphicx}
\usepackage{cite}
\usepackage{circuitikz} 
\usepackage[table]{xcolor} 
\usepackage{booktabs}
\usepackage{siunitx}
\usepackage{comment}

\begin{document}

\newcommand{\hdblarrow}{H\makebox[0.9ex][l]{$\downdownarrows$}-}
\title{{On-Chip Resonator for Nonlinear Kinetic Inductance Characterisation and Future Spectrometry Applications}}

\author{Patrick Ashworth$^{1,*}$, Peter S. Barry$^1$, Chris S. Benson$^1$, Harry Gordon-Moys$^1$, Izaak Morris$^1$
\thanks{$^1$School of Physics and Astronomy, Astronomy Instrumentation Group, Cardiff University, Cardiff, UK.\\
*Ashworthpd@Cardiff.ac.uk}
\thanks{Manuscript received August 1, 2025;}}

\markboth{Journal of \LaTeX\ Class Files,~Vol.~14, No.~8, August~2021}%
{Shell \MakeLowercase{\textit{et al.}}: A Sample Article Using IEEEtran.cls for IEEE Journals}
\maketitle

\begin{abstract} 
This work focuses on the development and demonstration of tunable superconducting on-chip resonator, leveraging the intrinsic current-dependent non-linear kinetic inductance of superconducting aluminium, and investigating the effect of oxygen content. Thin films are deposited using standard metal evaporation. We present results from a comprehensive study based on a series of evaporated Al thin films. This research aims to inform and constrain optimisation strategies for the design of mm-wave on-chip spectrometers, particularly regarding yield, resolution, and efficiency. By systematically varying film stoichiometry, we use a series of DC measurements to extract fundamental film properties such as resistivity, critical current and critical temperature. Furthermore, we employ low-loss DC-coupled microwave resonators to characterise both their microwave properties and the non-linear kinetic inductance, comparing these findings to a determined non-linear kinetic model. Finally, we discuss the possibility of usage in a parametric amplifier.  \end{abstract}

\section{Introduction}
Recent developments suggest that by utilising the non-linear kinetic inductance of a superconducting material, a DC bias can be used to control the central frequency of superconducting microwave resonator. This opens up new avenues for tune-ability~\cite{Vissers_2015}, enabling real-time tone tracking\cite{rouble2024demonstrationactivefeedbackcontrol}, dynamic feedback tuning~\cite{li2023fluxcoupledtunablesuperconducting}, and novel readout architectures. These capabilities are particularly useful in mitigating issues such as overlapping resonances, which often limit the yield and spectral resolution of current kinetic inductance detector (KID) arrays.

Another significant application for this research, is Superconducting travelling-wave parametric amplifiers (TWPAs). These low-noise cryogenic amplifiers, built upon thin-film superconducting transmission lines~\cite{eom2012widebandlownoisesuperconductingamplifier}, have garnered substantial interest over the past decade~\cite{Chaudhuri2017BroadbandPA,malnou2021performancekineticinductancetravelingwaveparametric,giachero2024kinetic}. Their appeal stems from their ability to deliver high gain and broadband amplification with noise performance approaching the quantum limit~\cite{PRXQuantum.2.010302}. In particular, Kinetic inductance travelling-wave parametric amplifiers (KI-TWPAs) ~\cite{ho2012wideband,malnou2022performance,giachero2024kinetic} offer higher saturation powers, determined by the critical currents of the materials. This characteristic makes them suitable for applications such as reading out large detector or qubit arrays~\cite{Castellanos_Beltran_2025}. 

The functionality of these devices is determined by the degree of non-linearity in their kinetic inductance. Optimising this functionality  requires reliable methods to characterise non-linear response of superconducting thin films. Several approaches have been proposed to tune this parameter including flux coupling~\cite{li2023fluxcoupledtunablesuperconducting}, Josephson junction-based architectures~\cite{PhysRevApplied.19.034021}, and Flux Qubit inductively coupled resonators~\cite{Chang_2023}. To probe this material property, we designed an on-chip coplanar waveguide resonator that can be directly tuned with a DC bias. Using a model from the Taylor expansion of the Ginzburg-Landau theory, Eq.~\eqref{eq:nonLinIstar}, we can directly extract the characteristic current scales ($I_\ast$) that define the quadratic and higher-order terms of the nonlinearity. Compared with previous methods, our approach provides a straightforward, reproducible measurement that is compatible with standard thin-film architectures.

\section{Theory}
Planar resonators provide a convenient method for probing the non-linear kinetic inductance due to their ease of fabrication and straightforward integration into superconducting circuits. A common implementation is the coplanar waveguide (CPW), consisting of a central signal line flanked by two ground planes.   

When an electromagnetic wave propagates along a CPW, it is reflected at a point of an impedance discontinuity. The resonant frequency of a CPW resonator is determined by the resonator length, $l$, between two impedance discontinuities, and is given by  
\begin{equation}
    f_{0n} = \frac{n c}{2 l \sqrt{\varepsilon_{\mathrm{eff}}}},
\end{equation}
where $n$ is the mode number, $c$ is the speed of light in vacuum, and $\varepsilon_{\mathrm{eff}}$ is the effective dielectric constant of the waveguide. 

A common method of implementing a impedance discontinuity is by introducing a capacitive gap in the transmission line~\cite{Bothner_2013}. This structure can be modelled as a lumped-element series capacitance, resulting in a resonator that is capacitively coupled at both ends.   Alternatively, a complementary impedance discontinuity can be created by shorting the three conductors of a CPW, which  introduces a shunt inductance $L_c$ between the centre conductor and ground of the CPW line~\cite{Matthaei1980MicrowaveFI}. Coupling via an effective capacitance relies on the electric field across a gap, while inductive coupling is mediated by magnetic flux between current loops~\cite{Gladchenko_2011,G_ppl_2008}.

The resonant frequency can also be approximated by an effective total inductance ($L_t$) and capacitance ($C$) of the resonator as 
\begin{equation}
   \omega_0 \approx 1/\sqrt{L_tC}
\end{equation}
where the total inductance comprises a geometric component ($L_g$), that is fixed for a given resonator geometry, and a kinetic component ($L_k$), which arises from the inertia of Cooper pairs in a superconducting film. At higher currents the value of $L_k$ becomes dependent on the applied current, modifying the resonant frequency as
\begin{equation}
\omega_0(I) = \frac{1}{\sqrt{L(I) C}} = \frac{1}{\sqrt{(L_g + L_k(I)) C}},
\end{equation}
where $L_k(I)$ now takes on an explicit current dependence.

The kinetic inductance is inherently non-linear and changes with the magnitude of the current. Derived from Ginzburg–Landau theory \cite{ginzburg2009theory},  $L_k (I)$ can be expressed as a power series as
\begin{equation}
L_{k}(I) = L_{k,0}\left( 1 + \frac{I^{2}}{I_{2*}^2} + \frac{I^{4}}{I_{4*}^4} + \dots \right),
\label{eq:nonLinIstar}
\end{equation}
where $L_{k,0}$ is the inductance per unit length in the absence of current. $I_{2*}$ and $I_{4*}$ represent the characteristic current scales for the quadratic and quartic orders of non-linearity, respectively \cite{eom2012widebandlownoisesuperconductingamplifier,Superconducting_Microresonators:Physics&Applications,2017PhDT.......472K, article}. Odd-order dependencies on current are absent because, for any infinitesimal segment of a superconducting transmission line, the kinetic inductance remains identical whether a current $+I$ or $-I$ is applied. 

The physical origin of this non-linear behaviour can be understood by considering the effect of a DC bias current. This bias not only breaks the $\pm I$ symmetry of the system but also influences the quasiparticle population in the superconductor. At non-zero DC currents, energy is injected into the system, which changes the ratio of quasiparticles to Cooper pairs, thereby modifying the kinetic inductance. Since $L_k$ is directly related to the superfluid density of Cooper pairs, a reduction in this density leads to an increase in kinetic inductance. Therefore, its resonance frequency, can be dynamically tuned.

For the equilibrium super-current non-linearity, in the limit of thin and narrow films, the scaling terms in
the kinetic inductance nonlinearity, $I_{2\ast} \text{ and } \text{I}_{\ast,4}$ in Eq.~\ref{eq:nonLinIstar}, can be related to the critical current $I_c$~\cite{Zhao_2022} by 
\begin{equation}
\label{characteristic current from the critical current}
    I_{2*} = \frac{3 \sqrt{3}}{2} I_c 
\end{equation}

\begin{equation}
     I_{4*} = \frac{1}{\sqrt[4]{3}} I_{2*}.
\end{equation}
By introducing a DC bias, the total current can be expressed as $I \rightarrow I_\text{DC} + I_\text{RF}$ where $I_\text{DC}$ is the bias current and $I_\text{RF}$ is the current component supplied by the input frequency.
Under this bias, the kinetic inductance also becomes dependent on the odd powers of the $I_\text{RF}$.

Since the resonant frequency scales as $1/\sqrt{L(I)C}$, it can be continuously tuned by adjusting the current up to its critical limit. The responsivity can be defined as $d|x|/dI$, where $x=\delta f_r/f_r=-\delta L/2L$ is the fractional frequency shift. To leading order, $d|x|/dI\approx I/I^2_{2*}$. This shows that responsivity can be enhanced either by choosing a material with a lower intrinsic $I_{2*}$  or by applying a larger bias current $I$.\\

\section{Design and Structure}

In order to probe the intrinsic non-linearity, a DC current needs to be supplied to the resonator trace. This is not possible with a standard capacitively-coupled lumped-element resonator as the resonator is electrically isolated.  While other DC biased inductively coupled lumped-element designs~\cite{Vissers_2015} have achieved this, they generally integrate explicit DC bias networks or flux-bias elements ~\cite{li2023fluxcoupledtunablesuperconducting} to control frequency, often requiring multiple materials.

Inspiration was taken from Bothner et al.~\cite{Bothner_2013}. An inductively coupled coplanar waveguide (CPW) was used as the base structure, as it allows a direct electrical connection to the centre trace. However, without a proper ground reference the RF signal is degraded, and simply connecting a DC input to the CPW would cause it to short to ground immediately. 
This issue was solved by introducing a large capacitive plate to ground to act as a DC filter. As can be seen in figure 4 the plane was split in two, preventing the current from being shunted around the resonator through the ground plane 

The design was initially evaluated in the Quite Universal Circuit Simulator (QUCS)~\cite{qucs} to verify inductive coupling with a capacitive path to ground. Once the design was proved feasible, simulations were performed in Sonnet EM, a method of moments EM simulation software that is capable of analysing RF structures. 
Through this software, we were able to design the architecture with the appropriate materials, providing a more accurate simulation of the performance of our design. 
\begin{figure}[H]
    \centering
    \includegraphics[width=1\linewidth]{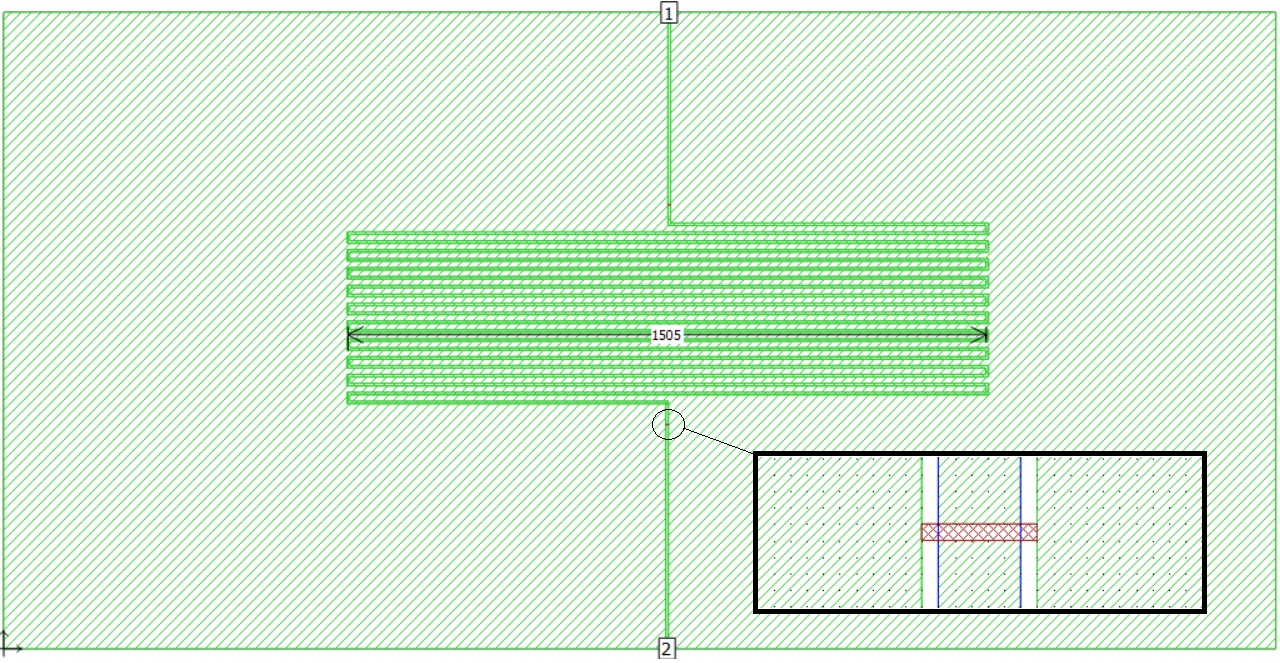}
    \caption{Sonnet Simulation and recreation of Bothner et al CPW}
    \label{Sonnet Simulation and recreation of Bothner et al CPW}
\end{figure} 

Fig.~\ref{Sonnet Simulation and recreation of Bothner et al CPW} shows an example layout of an inductively shorted half-wave resonator. Using Sonnet, we were able to tune the inductive short in a way that allowed us to cover a useful range of coupling strengths. 
\begin{figure}[H]
    \centering
    \includegraphics[width=1\linewidth]{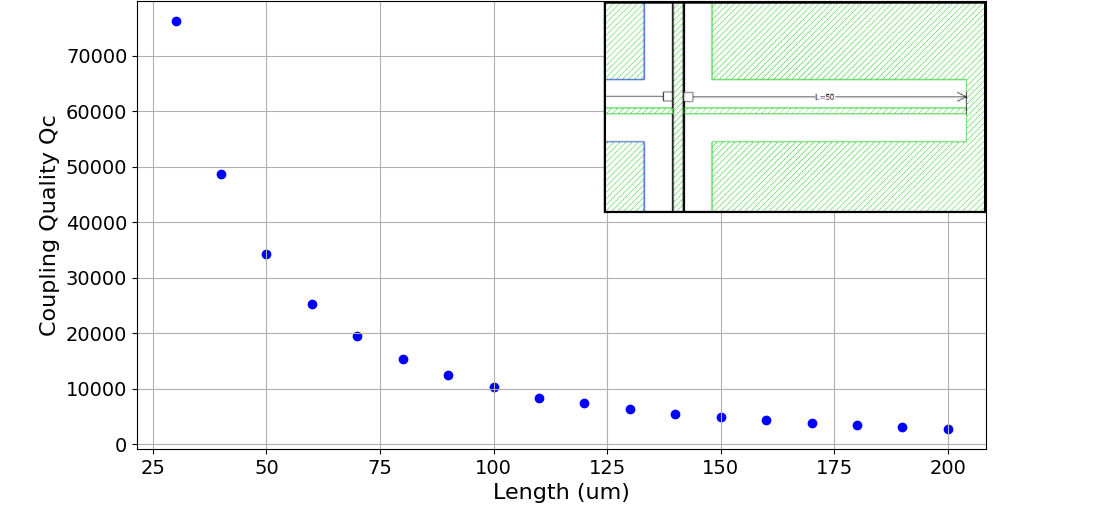}
    \caption{Sonnet simulation results for determining $Q_c$.}
    \label{Sonnet simulation results for determining Qc}
\end{figure}

The coupling quality factor ($Q_c$) was extracted by fitting a Lorentzian to the simulated $S_{21}$ response. The results, displayed in Fig.~\ref{Sonnet simulation results for determining Qc}, indicated that a smaller inductive short yields a higher $Q_c$. Based on this, a short length of \qty{50}{\micro\meter} was chosen to meet the target $Q_c$ of 10,000--50,000. This range is optimal for resonator readout, as it produces a resonator that does not require extremely high-resolution data acquisition while still allowing measurable shifts to be recorded.

An independent capacitively-coupled resonator was also included on the same device to verify that the resonance shift was not due to chip heating. A similar procedure was applied to the capacitively coupled resonator, where an interdigitated finger capacitor was used to target a coupling quality factor of approximately $35,000$ at a resonant frequency of \qty{1.7}{\giga\hertz}. In both cases, the final feedlines were adjusted to ensure proper matching to a \qty{50}{\ohm} load.

From basic circuit analysis and preliminary Sonnet simulations, we determined that a sufficiently large plate capacitor, combined with a thin-film dielectric, could function effectively as a DC filter. Due to the box layout, the available area was restricted to $1,100 \times 28,000$~\unit{\um}. Since the resonator transmission line did not occupy the entire chip, the remaining area was enough to implement the capacitor. A \SI{300}{\nano\meter} layer of silicon nitride was selected as the dielectric. 

\begin{figure}[]
    \centering
    \includegraphics[width=1\linewidth]{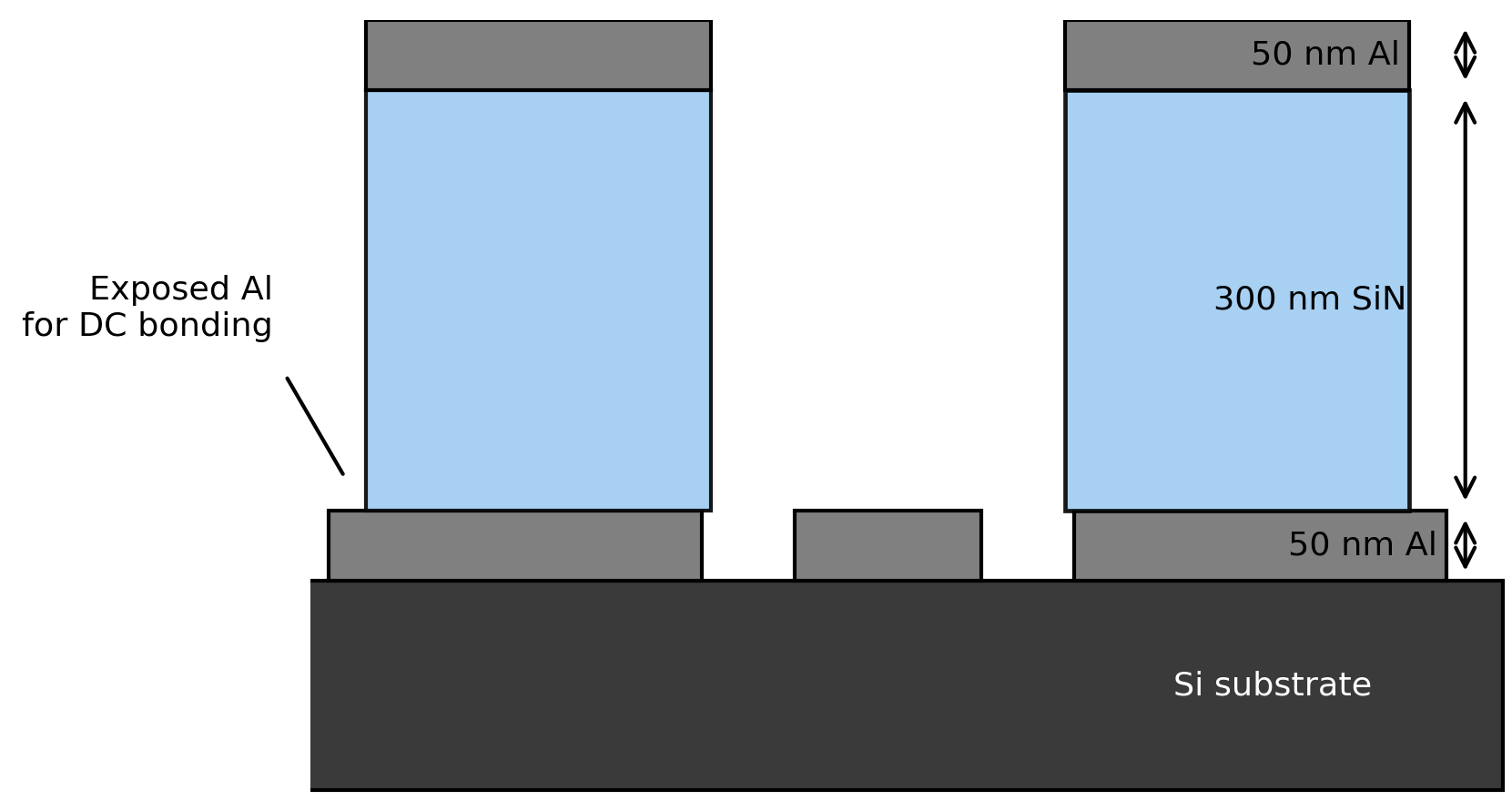}
    \caption{cross sectional diagram of on-chip structure}
    \label{cross sectional diagram of onchip structure}
\end{figure} 

\begin{figure}[ht]
    \centering
\includegraphics[width=1\linewidth]{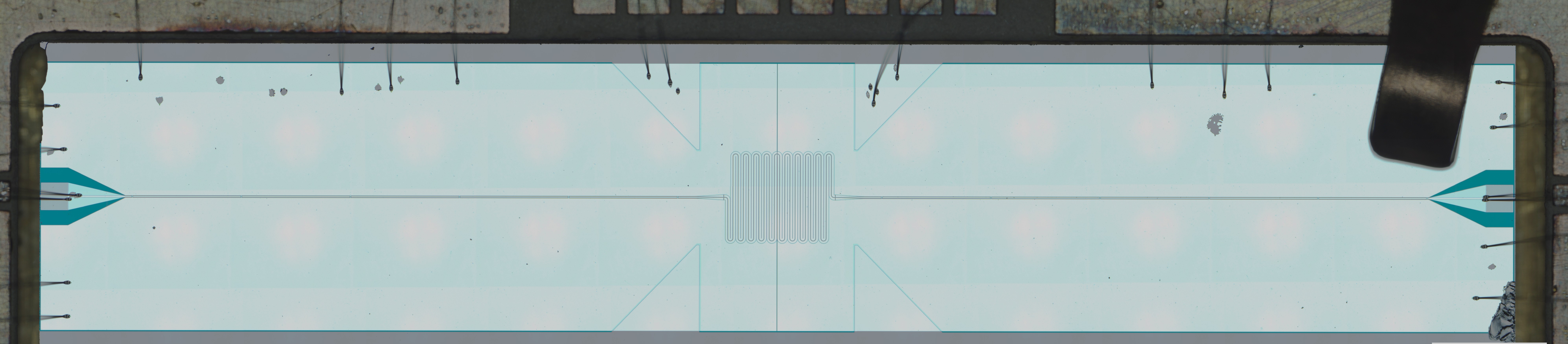}
    \caption{Final Fabricated design}
    \label{Final fabricated device in box}
\end{figure}

The final design choices were primarily dictated by the need to make electrical contact to each layer. Since there was no direct way to connect to the lower capacitor plane, which served as the DC input, two openings were patterned in the top ground plane to enable wire bonding to the lower plane. At the same time, care was needed to prevent an immediate short across the continuous lower plane. To address this, a narrow split was introduced in the layout, forcing the DC current to flow through the centre conductor. This modification can be seen in the final design Fig.~\ref{Final fabricated device in box}, where a section of the top ground plane was removed to provide access for bonding to the lower plane. Through our simulations we did not find that the large structures nor the split ground plane had a significant impacted our measurements.

\section{Fabrication}
The device fabrication consisted of sequential deposition, photolithographic patterning, wet etching, and lift-off steps to realise a three-layer aluminium structure separated by a silicon nitride dielectric. A schematic cross section of the device is shown in Fig.~\ref{cross sectional diagram of onchip structure}.

First, a bottom aluminium layer \qty{50}{\nano\meter} thick was deposited using an electron-beam sputter deposition system under high vacuum. A photolithographic process employing $AZ2020$ negative photoresist patterned with a Heidelberg MLA150 maskless lithography system. The aluminium was then wet-etched using a phosphoric-nitric-acetic acid mixture, forming the inductively coupled resonator pattern on the bottom layer.

A second lithography step was then performed to define the dielectric layer. The negative photoresist $AZ2035$ was patterned to create the liftoff mask, followed by the deposition of a \qty{300}{\nano\meter}-thick silicon nitride layer by reactive sputtering using a Syrus LC III system. The unwanted dielectric and resist were subsequently removed through a standard solvent-based lift-off process, leaving the patterned dielectric layer.

The upper capacitor plates were defined to provide an electrical connection to ground, while incorporating an opening to allow DC access to the bottom layer. The negative photoresist $AZ2035$ was patterned, followed by the deposition of a \qty{70}{\nano\meter}-thick aluminium layer. This metal layer was patterned by lift-off, completing the device structure.

\begin{figure}[H]
    \centering    \includegraphics[width=1\linewidth]{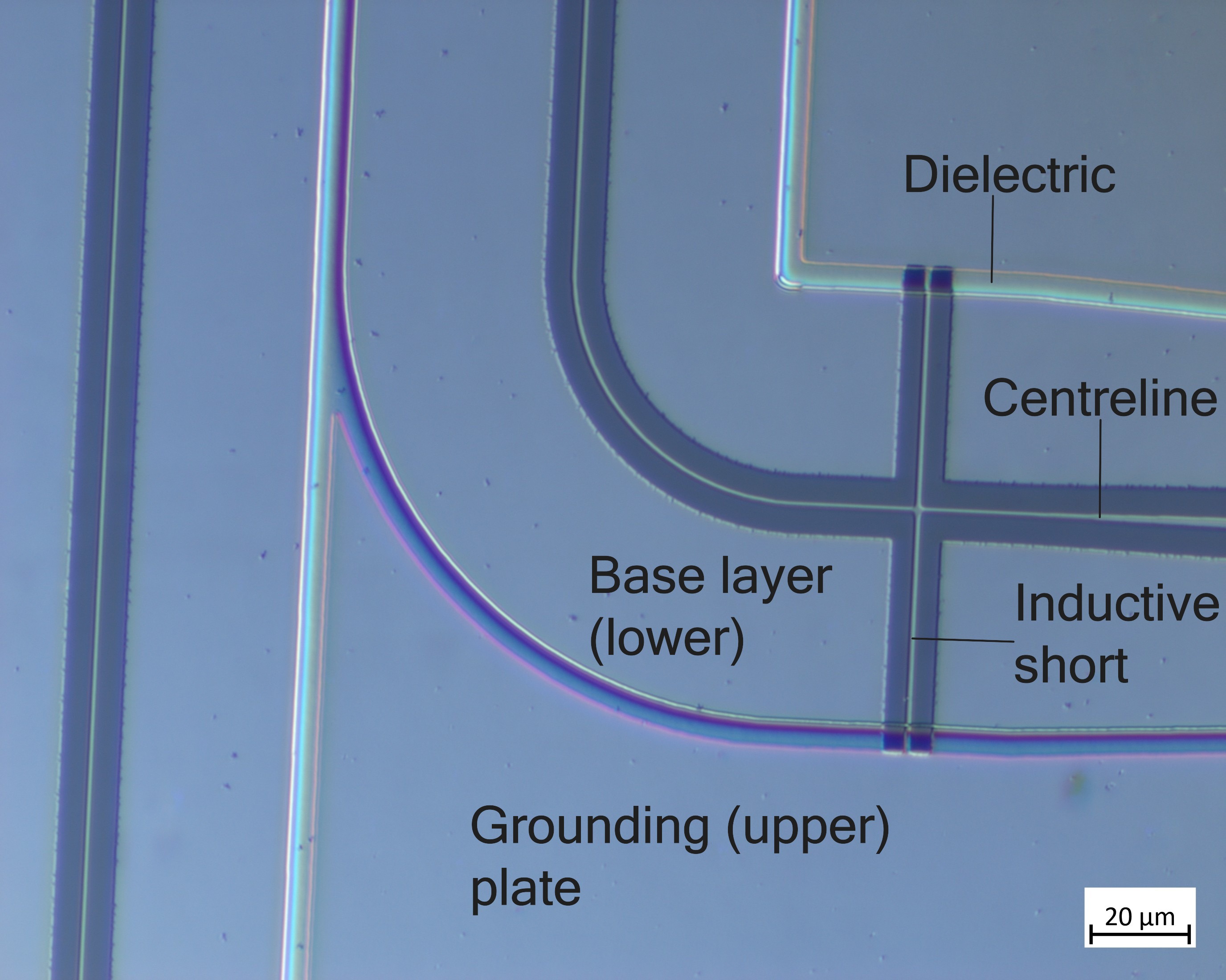}
    \caption{Microscope image of the fabricated inductive short.}
    \label{Microscope image of the fabricated inductive short.}
\end{figure}

Two wafers were fabricated, each patterned with an identical set of eight inductively coupled and eight capacitively coupled resonators. As part of the deposition, one wafer underwent a titanium gettering process prior to deposition. In this procedure, titanium is evaporated in the vacuum chamber to reduce residual oxygen. The substrate shutter remains closed to prevent titanium deposition on to the sample. For the non-gettered sample, a pre deposition base pressure of \SI{4.6e-7}{\milli\bar} was reached, whereas the gettering process yielded a reduced pressure of \SI{3.6e-8}{\milli\bar}. Through gettering, we expect to produce an aluminium film with reduced oxygen content and fewer defects. 
Fig.~\ref{Microscope image of the fabricated inductive short.} shows the patterned inductive short line.

\section{Experimental Setup}

\begin{figure}
    \centering
    \includegraphics[width=0.8\linewidth]{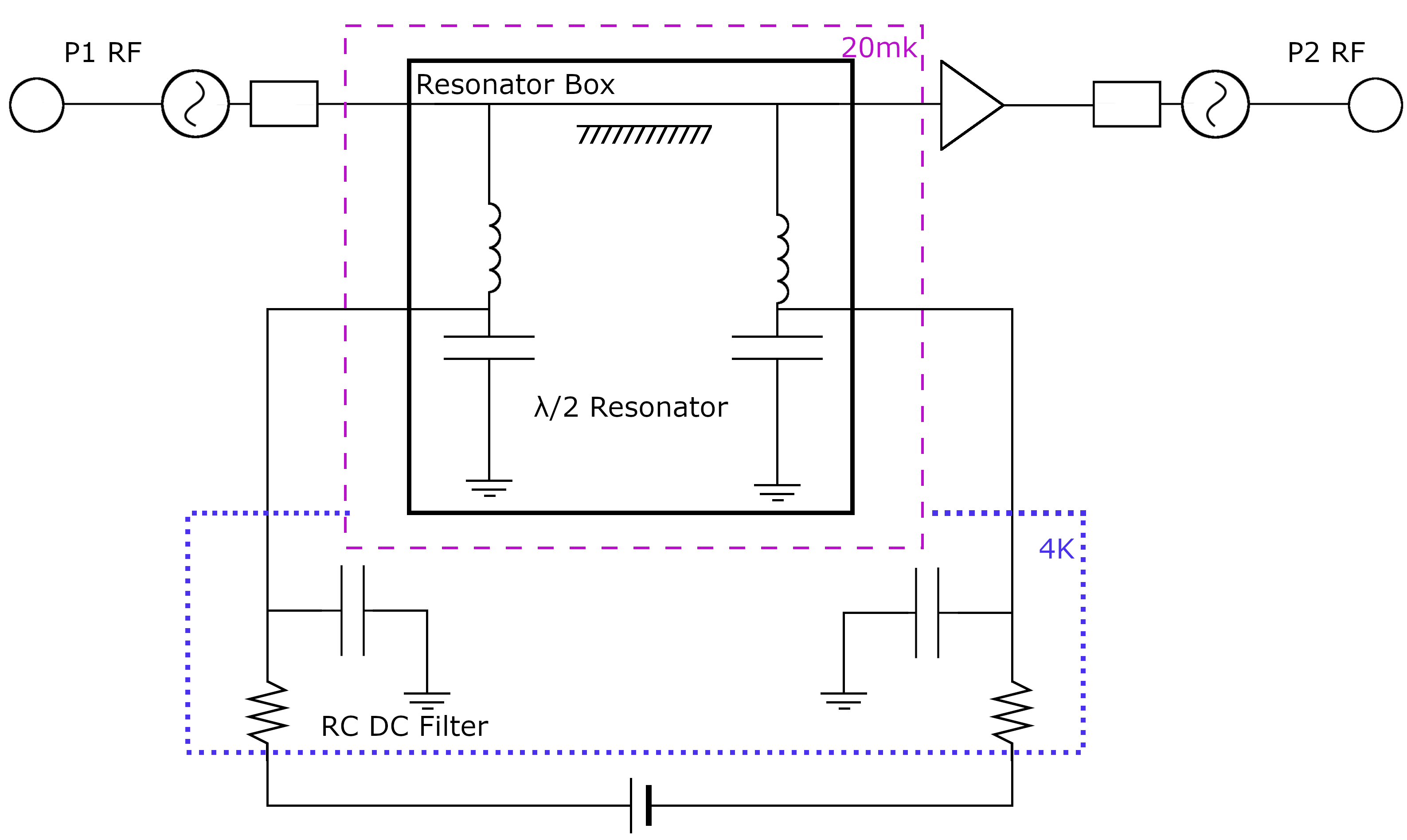}
    \caption{Resonator box and Biasing configuration circuit}
    \label{Resonaotr box & Biasing configuration circuit}
\end{figure}

The sample was wirebonded to standard \qty{50}{\ohm} SMA connectors, allowing microwave readout through the transmission line. The resonators were mounted inside a shielded Au-plated Al box equipped with a coaxial RF input/output line and a two-pin DC input, then installed onto the 20~mK stage of a BlueFors cryostat (model LD250). The DC line was routed through an RC low-pass filter in series with an Eccosorb powder filter to suppress high-frequency noise and prevent unwanted excitation of the resonator \cite{phdthesis}. A circuit diagram of the configuration is shown in Fig.~\ref{Resonaotr box & Biasing configuration circuit}.

A vector network analyser (VNA) was used to measure the forward transmission parameter, $S_{21}$, enabling tracking of the resonant frequency as a function of applied DC bias and temperature. A DC current was applied in discrete steps, with measurements taken at each value, until the resonator response indicated the onset of a transition, which we attribute to exceeding the critical current. Initial sweeps were performed until this transition occurred, with five repetitions recorded for each scan. Sweeps at approximately \qty{24}{\milli\kelvin} were extracted, and each current sweep was averaged before plotting.

\section{Results and Analysis}

\begin{figure}[H]
\centering
\includegraphics[width=1\linewidth]{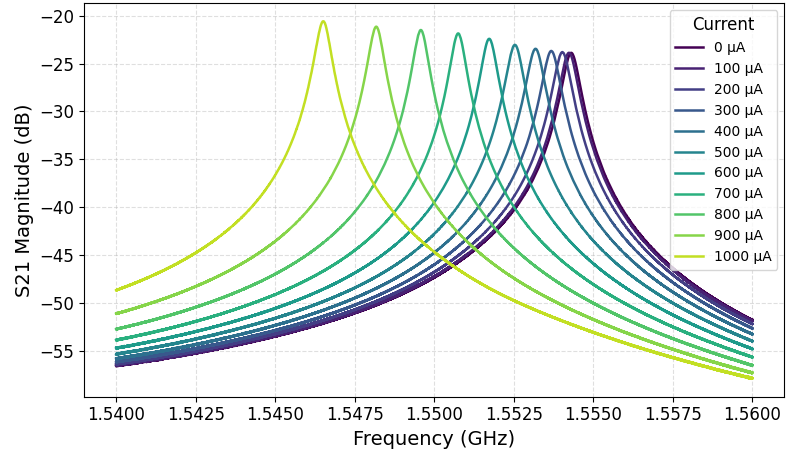}
\caption{S21 data for Gettered Al at bias DC}
\label{S21_data_for_Gettered_Al_at_bias_Dc}
\end{figure}

Fig.~\ref{S21_data_for_Gettered_Al_at_bias_Dc} shows the resonant frequency as a function of applied bias for the gettered sample, which is representative of the sweep data from the non-gettered sample.

A key parameter in understanding the current-dependent frequency shift is the kinetic inductance fraction, $\alpha_k$, which quantifies the contribution of the kinetic inductance to the total inductance of the system. Determining $\alpha_k$ requires independent measurements of both the geometric inductance, $L_g$, and the zero-current kinetic inductance, $L_{k,0}$. In this work, we employed two independent methods to extract $\alpha_k$.

Letting $L_{k,0}$ denote the zero-current kinetic inductance, and $I_{n*}$ the characteristic current scales of the non-linearity, we define a fractional frequency shift as
\begin{equation}
\frac{\delta \omega}{\omega_0} \equiv \left(\frac{f_{I}}{f_{0}}\right)^2 - 1 = \frac{L(0)}{L(I)} - 1,
\end{equation}
and the kinetic inductance fraction as
\begin{equation}
\alpha_k = \frac{L_{k,0}}{L_{k,0} + L_g}.
\label{eq:alphak}
\end{equation}
%
Substituting the expressions for the total inductance at zero current, $L(0) = L_g + L_{k,0}$, and at finite current, $L(I)$, the relative frequency shift can be expressed as

\begin{align} \frac{\delta \omega}{\omega_0} &= \frac{L_{k,0}/\alpha_k}{L_{k,0}/\alpha_k + L_{k,0} \epsilon} - 1 \\ &= \frac{1}{1 + \alpha_k \left( \frac{I^2}{I_{2*}^2} + \frac{I^4}{I_{4*}^4} + \cdots \right)} - 1. \end{align}

\subsection{Method 1: Component-Based Calculation} 
The value for $L_g$, and therefore $\alpha_k$, can be estimated from EM simulation. By simulating the resonator and manually setting the material's kinetic inductance to a known value ($L_{\text{add}}$), the resulting frequency shift can be used to extract the purely geometric component using the following relation

\begin{equation} L_g = L_{\text{add}} \, \frac{\omega_{0,x}^{2}}{\omega_{0,0}^{2} - \omega_{0,x}^{2}}. \end{equation}

The zero-current kinetic inductance, $L_{k,0}$, was calculated from material properties measured in a Physical Property Measurement System (PPMS). This cryogenic platform provides precise temperature control and enables characterisation of the critical temperature and film resistivity. 
$L_{k,0}$ is derived from the normal state sheet resistance ($R_s$) and the superconducting critical temperature ($T_c$) as

\begin{equation} L_k = \frac{\hbar R_s}{\pi \Delta_0}, \quad \text{where} \quad \Delta_0 \approx 1.76 \, k_{\mathrm{B}} T_c. \end{equation} 

With these two components, $\alpha_k$ is calculated directly using its definition in Eq.~\ref{eq:alphak}.

\subsection
{Method 2: Temperature-Dependent Fitting}~
The second method determines $\alpha_k$ by analysing the resonator's response to temperature, using the Mattis--Bardeen (MB) formalism for quasiparticle-induced complex conductivity. The model for the fractional frequency shift as a function of temperature is given by \cite{zmuidzinas2012superconducting}
\begin{equation} 
x(T) = - \frac{\alpha_k}{4 \Delta N_0} \, n_{\text{qp}}(T, \Delta) \, S_2(f_0, T, \Delta), 
\end{equation} 
where the components are the thermal quasiparticle density, $n_{\text{qp}}$, and the MB conductivity kernel, $S_2$. 
The quasiparticle density is expressed as
\[ n_{\text{qp}}(T, \Delta) = 2 N_0 \sqrt{2 \pi k_B T \Delta} \, \exp\!\left(-\frac{\Delta}{k_B T}\right), \] 
a low-temperature approximation (T $<<$ Tc) obtained by assuming a Boltzmann form for the quasiparticle occupation rather than the full Fermi–Dirac distribution. The kernel $S2$ given by 
\[ S_2(f_0, T, \Delta) = 1 + \sqrt{\frac{2\Delta}{\pi k_B T}} \, I_0\!\left(\xi\right) \, e^{-\xi}, \qquad \xi = \frac{\hbar \pi f_0}{k_B T}, \]
with $N_0$ being the single-spin density of states, $\Delta$ the superconducting energy gap, and $I_0$ the zeroth-order modified Bessel function. 
To apply this model, the resonance frequency was measured as a function of temperature at zero DC bias. The experimental fractional frequency shift was computed as $x(T) = f(T)/f_0 - 1$. A non-linear least-squares fit was then performed between the experimental data and the theoretical model for $x(T)$ using the \texttt{lmfit} \cite{newville2016lmfit} library. In this fit, the kinetic inductance fraction ($\alpha_k$) and the critical temperature ($T_c$) were treated as free parameters. The single-spin density of states was fixed at $N_{0} = 1.72 \times 10^{10}\,\unit{\micro\meter^{-3}\electronvolt^{-1}}$~\cite{moshel2025propagationvelocitymeasurementssubstrate}. 
This allowed $\alpha_k$ and $T_c$ to be extracted directly from the temperature-dependent data. The values obtained using these two methods are shown in Table~\ref{Determined values from PPMS and fitting}.

\begin{figure}[H]
    \centering
    \includegraphics[width=1\linewidth]{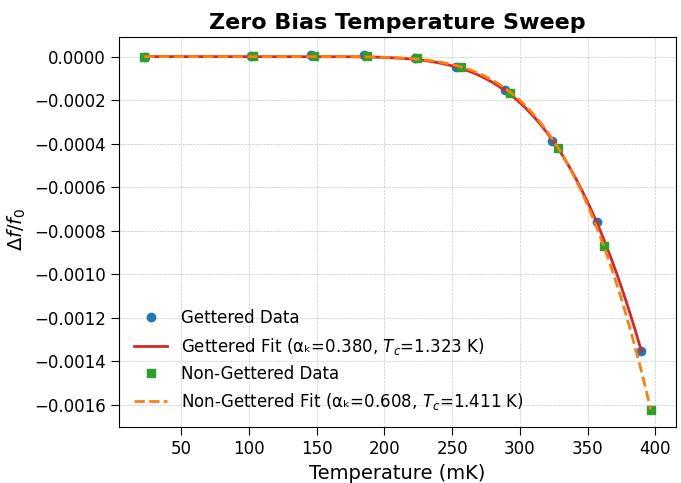}
    \caption{Mattis-Bardeen fit to temperature sweep data at zero DC bias}
    \label{Mattis-Bardeen fit to temperature sweep data at 0 DC}
\end{figure}

Fig.~\ref{Mattis-Bardeen fit to temperature sweep data at 0 DC} displays the MB fit for a range of temperatures at zero DC, for both samples.

\begin{figure}[H]
    \centering
    \includegraphics[width=1\linewidth]{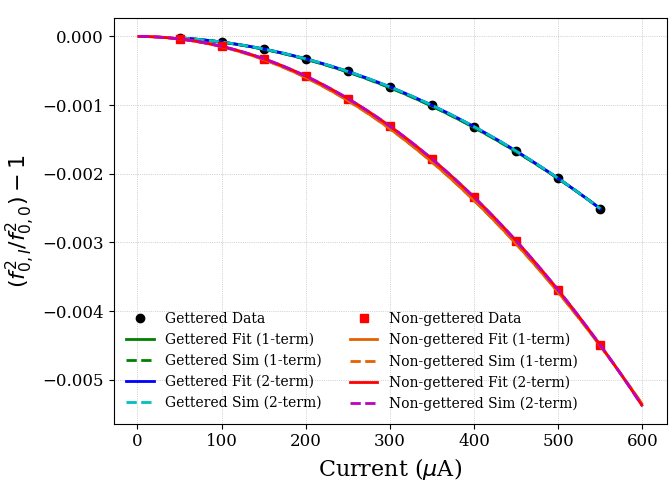}
    \caption{Extraction of $I_{n*}$ in Al Resonators with and without Ti Gettering. 
    Al Gettered Data; Fit 1-term ($I_{2*} = 6760.8$); Sim 1-term ($I_{2*} = 4672.3$); 
    Fit 2-term ($I_{2*} = 6840.3, I_{4*} = 4838.8$); Sim 2-term ($I_{2*} = 4727.3, I_{4*} = 4022.6$); 
    Non-gettered Data; Fit 1-term ($I_{2*} = 6376.8$); Sim 1-term ($I_{2*} = 4002.0$); 
    Fit 2-term ($I_{2*} = 6508, I_{4*} = 4260$); Sim 2-term ($I_{2*} = 4084.3, I_{4*}= 3374.8$). }
    \label{fig: I*_extraction_plot}
\end{figure}

The fractional frequency shift as a function of DC bias current for both the gettered and non-gettered resonators is shown in Fig.~\ref{fig: I*_extraction_plot}. Before discussing the fits, we can make a simple estimate of the characteristic current $I_{2*}$ from the critical current $I_c$ using Eq.~\ref{characteristic current from the critical current}. The range of data seen on the x-axis of Fig.~\ref{fig: I*_extraction_plot}, is determined by the onset of a change in the resonator transmission. This change we attribute to exceeding the critical current, allowing approximate values of $I_c$ to be determined. For the gettered sample, $I_c \approx$ \qty{0.55}{\milli A}, corresponding to $I_{2*}\approx$ \qty{1.43}{\milli A}. For the non-gettered sample, $I_c \approx$ \qty{0.6}{{\mA}}, corresponding to an $I_{2*}\approx$ \qty{1.56}{\milli A}. These estimates suggest, the non-gettered film should have a higher value of $I_{2*}$.

\begin{table}[h!]
\centering
\rowcolors{2}{cyan!10}{white} 
\begin{tabular}{lcc}
\toprule
\textbf{Terms} & \textbf{Gettered} & \textbf{NonGettered} \\
\midrule
$R_s$ (\unit{\ohm}/sq)     & 0.2916 & 0.4282 \\
$T_c$ (\unit{\kelvin})     & 1.274  & 1.314 \\
$L_k$/sq (pH/sq)           & 0.316  & 0.45 \\
Sim $L_g$ (pH)             & 1.428  & 1.428 \\
Sim $\alpha_k$             & 0.18   & 0.239 \\
Fit $\alpha_k$             & 0.380  & 0.608 \\
Fit $T_c$ (\unit{\kelvin}) & 1.323   & 1.411 \\
2 term Fit (\unit{\uA})    & 6840.3 $I_{2*}$, 4838.8 $I_{4*}$ & 6508 $I_{2*}$, 4260 $I_{4*}$ \\
2 term Sim (\unit{\uA})    & 4727.3 $I_{2*}$, 4022.6 $I_{4*}$ & 4084.3 $I_{2*}$, 3374.8 $I_{4*}$ \\
Sim $I_{*2,0}$ (\unit{\mA})& 11.14  & 8.35 \\
Fit $I_{*2,0}$ (\unit{\mA})&  11.09  & 8.34  \\
\bottomrule
\end{tabular}
\vspace{4pt}
\caption{Measured values from PPMS and model fitting}
\label{Determined values from PPMS and fitting}
\end{table}

Fig.~\ref{fig: I*_extraction_plot} confirms that the non-gettered aluminium film exhibits a more pronounced non-linear response, displaying a larger fractional frequency shift for a given bias current, suggesting a smaller $I_{2*}$ value. 

The data was fit using both a single-term (quadratic) and a two-term (quadratic + quartic) model for the kinetic inductance non-linearity of the form, 
\begin{equation}
y(I) = \frac{1}{1 + \alpha_k \left[ \left(\frac{I}{I_{2*}}\right)^{2} + \left(\frac{I}{I_{4*}}\right)^{4} \right]} - 1 .
\label{nonlinear_kinetic_inductance_frequency_shift_formula}
\end{equation}

The values of $\alpha_k$ differ in magnitude between samples for both methods. Although this variation was anticipated due to differences in film quality, it raises a subtlety when interpreting the estimates of $I_{2*}$ as extracted from the fit. The intrinsic value of $I_{*2,0}$ can be extracted by scaling the measured $I_{*2}$ by $1/\sqrt{\alpha_{k}}$.

The values of $I_{*2,0}$ are included in Table 1, and show good agreement for the two methods of estimating $\alpha_k$. While the source of the discrepancy between the estimated and measured magnitude of $I_{*2}$ from Eq.~\ref{characteristic current from the critical current} is not yet fully understood, the ratio of the values for the two films appears to be reasonably consistent (1.3 vs 1.5) given the rudimentary estimate from $I_c$.

\section{Conclusion}
A superconducting resonator with an integrated DC input line was successfully designed, fabricated, and characterized. It exhibited a repeatable, non-linear resonance frequency shift as a function of the applied DC current, with measurements on a capacitive reference resonator confirming that the observed shifts were not due to heating.

Comparison of gettered (prepared with Ti sputtering while the substrate shutter was closed to reduce oxygen content) and non-gettered (untreated) aluminium films revealed that the non-gettered resonator exhibited a reduced nonlinearity, consistent with increased disorder and impurity scattering~\cite{Deshpande_2025}. These results indicate that film quality plays a significant role in determining the non-linear behaviour of superconducting resonators under DC bias.

Direct Mattis--Bardeen fits indicate that the non-gettered aluminium resonator has a substantially higher kinetic inductance fraction, $\alpha_k = 0.608 \pm 0.01\ (3.02\%)$, compared with $\alpha_k = 0.379 \pm 0.01\ (4.95\%)$ for the gettered film, while the corresponding critical temperatures are $T_c = 1.41 \pm 0.006\ \mathrm{K}\ (0.45\%)$ and $T_c = 1.323 \pm 0.021\ \mathrm{K}\ (0.77\%)$, respectively. Both datasets are well described by the Mattis--Bardeen model ($R^2>0.999$) with exceptionally low reduced chi-square values. The larger $\alpha_k$ in the non-gettered film is consistent with enhanced disorder or impurity-related kinetic inductance, whereas the gettered film shows reduced kinetic inductance and a modest $T_c$ shift, reflecting the measurable impact of the gettering process on superconducting film properties.

\bibliographystyle{IEEEtran}
\bibliography{DC_Bias_Resonator}

\end{document}